\documentclass[12pt]{article}
\usepackage{amssymb,amsmath,amscd,amsthm}

\newtheorem{theorem}{Theorem}[section]

\newtheorem{lemma}[theorem]{Lemma}

\font\bbc=msbm10 scaled 1200

\newcommand{\R}{\mbox {\bbc R}}

\newcommand{\sol}{u^{\text{sc}}}
\newcommand{\obs}{\mathcal S}

\date{}

\begin{document}

\author{ Evgeny Lakshtanov\thanks{Department of Mathematics, Aveiro University, Aveiro 3810, Portugal.  This work was supported by Portuguese funds through the CIDMA - Center for Research and Development in Mathematics and Applications and the Portuguese Foundation for Science and Technology (``FCT--Fund\c{c}\~{a}o para a Ci\^{e}ncia e a Tecnologia''), within project UID/MAT/0416/2013 (lakshtanov@ua.pt).} \and
 Boris Vainberg\thanks{Department
of Mathematics and Statistics, University of North Carolina,
Charlotte, NC 28223, USA. The work was partially supported  by the NSF grant DMS-1008132 (brvainbe@uncc.edu).}}

\title{Uniqueness in potential scattering with reduced data }

 \maketitle
\begin{abstract}
We consider inverse potential scattering problems where the source of the incident waves is located on a smooth closed surface outside of the inhomogeneity of the media. The scattered waves are measured on the same surface at a fixed value of the energy. We show that this data determines the bounded potential uniquely.
\end{abstract}


\section{Introduction}

Consider a potential scattering problem
\begin{equation}\label{eq2}
-\Delta \sol -\lambda n(x)\sol =\lambda [n(x)-1]u^{inc}, \quad x \in \mathbb R^d, ~ \lambda=k^2>0, \\
\end{equation}
where the support of $n(x)-1$ belongs to a bounded domain ${\mathcal O}$, $~n(x)$ is  uniformly bounded  in $\overline{{\mathcal O}}$, and the solution $\sol$ satisfies the radiation condition:
\begin{equation}\label{rc}
\sol=\sol_\infty(k,\theta)\frac{e^{ikr}}{r^{\frac{d-1}{2}}}  + O \left (r^{-\frac{d+1}{2}} \right ),\quad \theta=\frac{x}{r},~~r=|x|\to\infty.
\end{equation}
Here $u^{inc}$ is an incident wave that satisfies the Helmholtz equation in $\mathbb R^d \backslash \obs$  where $\obs$ is a set, where sources are  distributed. We assume that $\obs$ is a smooth surface that is a boundary of a bounded domain $B$ located outside of $\mathcal O$. To be more exact,
\begin{equation}\label{new}
u^{inc}(x)=\int_{\obs} \frac{e^{-ik|x-y|}}{|x-y|} \varphi(y) dS_y, \quad  \varphi\in L_2(\obs), \quad x\in \mathbb R^d.
\end{equation}

There are many results on recovering information on the scatterer from the backscattering data. For example, results on the uniqueness of the solution of the inverse problem can be found in \cite{er1},\cite{er2},\cite{er3},\cite{p},\cite{rakesh} and recovering of singularities was studied in \cite{small},\cite{paiv},\cite{ruiz}. In all the papers above, it was
assumed that the echo data are available for incident waves coming from all the directions.

There are important applications when an observer has an access to the support of the potential only from one side. Additionally, the incident waves can be often emitted only from a bounded region, and not from
infinitely remote points as in the classical backscattering problem. Recently, such a non-stationary potential scattering problem (with a potential that is smooth in $\R^3$) has been studied by Rakesh and Uhlmann \cite{rakesh2}.
They assumed that the incident waves were emitted from points $x$ varying in some sphere. They show the uniqueness for potentials with some restrictions on angular derivatives.
In \cite{lv2015} we considered the scattering problem (\ref{eq2}) when the incident waves were emitted from surface $\obs$ and the receivers are also distributed over the same surface $\obs$, i.e., the following data are available:
\begin{equation}\label{data15}
\left \{ \sol|_{\obs} : ~ u^{inc} \mbox{  emitted from  } \obs \right  \}.
\end{equation}
We have shown that data (\ref{data15}) allows one to determine the interior eigenvalues of the scatterer. In this article, we prove a uniqueness result. Namely,
let us fix $\lambda >0$ that is not a Dirichlet Laplacian eigenvalue for the domain $B$ bounded by $\obs$. We show that data (\ref{data15}) for a fixed value of $\lambda>0$ determines the potential $n(\cdot)$ uniquely. We also will assume that $\lambda$ is not an eigenvalue of the Dirichlet problem
for the equation $(-\Delta-\lambda n(x))u=0$ in $\mathcal O$. Since the support of $n$ is bounded, the latter requirement can be enforced by a slight extension of $\mathcal O$. Without loss of the generality, we can assume that the boundary of $\mathcal O$ is infinitely smooth and the support of $n(x)-1$ is located strictly inside of $\mathcal O$.

Note also that the problem we consider is different from the problem of recovering of the potential from partial Cauchy data (see e.g. \cite{bu}). In the latter problem, it is assumed that Cauchy data are available for all sufficiently regular solutions of the wave equation. The situation is different in the problem under consideration. Here only the fields on $\obs$ are known that are produced by waves emitted from $\obs$.

{\bf Acknowledgments.} The authors are thankful to Rakesh, Eemeli Bl\aa sten, Uwe K$\rm{\ddot{a}}$hler and Lassi P$\rm{\ddot{a}}$iv$\!\rm{\ddot{a}}$rinta, David Colton and Armin Lechleiter for useful discussions.
\section{The main result}
From now on, for the sake of simplicity of notations, we assume that $d=3$. Define operator
$$
\mathcal L ~ : ~ L_2(\obs) \rightarrow L_2(\partial \mathcal O), \quad \mathcal L^* ~ : ~ L_2(\partial \mathcal O)
\rightarrow L_2(\obs) ,
$$
\begin{equation}\label{opL}
(\mathcal L \varphi)(x)= \int_{\obs} \frac{e^{-ik|x-y|}}{|x-y|} \varphi(y) dS_y, \quad
(\mathcal L^* \mu)(x)= \int_{\partial \mathcal O} \frac{e^{ik|x-y|}}{|x-y|} \mu(y) dS_y, \quad k=\sqrt\lambda>0.
\end{equation}
\begin{lemma}\label{6DecD}
Suppose that $\lambda>0$ is not an eigenvalue of the negative Dirichlet Laplacian in either of the domains $\mathcal O$ or $B$ (with the boundary $\obs$).
Then operators $\mathcal L, \mathcal L^*$ have dense ranges.
\end{lemma}
{\bf Remark.} An outline of this proof can be found in \cite{lv2015}. Note also the integral kernels of operators $\mathcal L, \mathcal L^*$ are infinitely smooth, and the arguments below prove that their ranges are dense in any Sobolev space $H^s,~s\geq 0,$ not only in $L_2$.

{\bf Proof.}
Let us prove that the range of $\mathcal L$ is dense.
Obviously, it is enough to show that the kernel of the operator $\mathcal L^* $ is trivial. Assume that the opposite is true. Then there exists $\mu\in L_2(\partial \mathcal O)$ such that $\mu\not\equiv 0$
and function
$$
u:= \int_{\partial \mathcal O} \frac{e^{ik|x-y|}}{|x-y|} \mu(y) dS_{y}, \quad x\in \R^3,\quad k=\sqrt\lambda>0,
$$
which is defined on $\R^3$ and coincides with $\mathcal L^* \mu$ on $\obs$, vanishes on $\obs$. Since
\[
 (-\Delta-\lambda)u=0, \quad x\notin \partial \mathcal O,
\]
and $\lambda$ is not an eigenvalue of the Dirichlet problem in $B, ~u\equiv 0$ on $B$. Then from the equation above it follows that $u\equiv 0$ on $\R^3\setminus \mathcal O$.

If $\mu$ is continuous, the proof can be completed in a couple of lines using the potential theory. Indeed, $u$ is continuous in $\R^3$ in this case. Thus $u$ satisfies the Helmholtz equation and the homogeneous Dirichlet boundary condition in $\mathcal O$. Since $\lambda $ is not an eigenvalue, it follows that $u\equiv 0$ in $\mathcal O$, i.e., $u\equiv 0$ in $\R^3$. The latter contradicts the fact that the jump of the normal derivative of $u$ on $\partial\mathcal O$ is equal to $-4\pi\mu\not\equiv 0$.

If $\mu\in L_2(\partial \mathcal O)$, then we approximate $\mu$ in $L_2(\partial \mathcal O)$ by smooth functions $\mu_n$. Consider
\begin{equation}\label{unn}
u_n= \int_{\partial \mathcal O} \frac{e^{ik|x-y|}}{|x-y|} \mu_n(y) dS_{y}, \quad x\in \R^3.
\end{equation}
If we restrict $u_n$ to $\partial \mathcal O$, then operator (\ref{unn}) becomes a pseudo-differential operator on $\partial \mathcal O$ of order $-1$, and therefore $u_n|_{\partial \mathcal O}$ has a limit in $H^{1/2}(\partial \mathcal O)$ as $n\to \infty$ (as well as in $H^1(\partial \mathcal O)$). Functions $u_n$ satisfy the Helmholtz equation outside of $\partial \mathcal O$, and they satisfy the radiation conditions. Thus the convergence of $u_n|_{\partial \mathcal O}$ and standard a priori estimates in $H^1$ for the solutions of the Helmholtz equation imply that functions $u_n$ converge in $H^1( \mathcal O)$ and in $H^1_{loc}(\R^3\backslash \mathcal O)$. Obviously, they converge to $u\equiv 0$ in $H^1_{loc}(\R^3\backslash \mathcal O)$. Thus
\[
u_n|_{\partial \mathcal O}\to 0 \quad {\rm in}~~H^{1/2}(\partial \mathcal O)    \quad {\rm as} ~~ n\to \infty.
\]
Hence $u_n$ converge in $H^1( \mathcal O)$ to a solution of the homogeneous Dirichlet problem. Since $\lambda$ is not an eigenvalue of the Dirichlet problem in $\mathcal O$, this implies that $u_n\to 0$ in $H^1( \mathcal O)$.

Since $\mu_n$ is smooth, the jump on $\partial\mathcal O$ of the normal derivative of the potential $u_n$ defined by (\ref{unn}) is equal to $-4\pi\mu_n\not\equiv 0$. On the other hand, the normal derivatives of weak (in $H^1$) solutions of the Helmholtz equation are well defined, and from the weak (in $H^1$) convergence of $u_n$ to zero it follows that this jump (which is equal to $\mu_n$) tends to zero in $H^{-1/2}(\partial \mathcal O)$. Since $\mu_n$ approximates $\mu$ in $L_2(\partial \mathcal O)$, it follows that $\mu=0$. This contradicts the assumption made in the first lines of the proof.
Thus the density of the range of the operator $\mathcal L$ is proved. Similar arguments are valid for $\mathcal L^*$.

 \qed

{\bf Definition}. Consider the near-field operator
$$
F_S =F_S(\lambda) : ~ L_2(\obs)\rightarrow L_2(\obs), \quad F_S\varphi = \sol|_{\obs}, ~~\varphi\in L_2(\obs) ,
$$
where $\sol$ is the solution of (\ref{eq2}) with $u^{inc}$ given by (\ref{new}).

Note that formula (\ref{new}) represents waves coming to $\obs$, while waves emitted from $\obs$ have the different sign in the exponent. Thus $F_S\varphi$ is not the scattered wave produced by sources on $\obs $ with the density $\varphi$.  However, $\sol|_{\obs}=F_S\varphi$ can be obtained (and measured) as a scattered field on $\obs $ produced by some waves emitted from $\obs$. Namely, the following lemma holds (see \cite{lv2015}).
\begin{lemma}\label{11J1}
Suppose that $\lambda>0$ is not an eigenvalue of the negative Dirichlet Laplacian in either of the domains $\mathcal O$ or $B$. Then  for each $\varphi\in L_2(\obs)$, one can construct a sequence $\psi_n\in L_2(\obs)$ of the source densities such that $F_S\varphi=\lim_{n\to 0}\sol_n|_{\obs}$, where the limit is taken in the space $ L_2(\obs)$ and $\sol_n$ is the solution of (\ref{eq2})  with
$$
u^{inc}(x)=\int_{\obs} \frac{e^{ik|x-y|}}{|x-y|} \psi_n(y) dS_y, \quad  \psi_n\in L_2(\obs), \quad x\in \mathbb R^d.
$$
 One can determine the source densities $\psi_n$ without a priori knowledge of $\mathcal O$ except a value of an $\varepsilon>0$ such that $\mathcal O$ is located inside of the ball $|x|<1/\varepsilon$.
\end{lemma}
{\bf Proof.} Consider a bounded domain $\widetilde{\mathcal O}$ that contains $\overline{\mathcal O}$  and  such that ${\rm dist}(B, \widetilde{\mathcal O})>0$. For example, one can take $\widetilde{\mathcal O}=(\mathbb R^d\backslash\overline{ B_\varepsilon})\bigcap\{|x|<1/\varepsilon\}$, where
$B_\varepsilon$ is the $\varepsilon$-extension of $B$ and $\varepsilon>0$ is small enough. Without loss of generality, we can assume that the boundary of $\widetilde{\mathcal O}$ is infinitely smooth and $\lambda$ is not an eigenvalue of the negative Dirichlet Laplacian in $\widetilde{\mathcal O}$.

From Lemma \ref{6DecD} it follows that the range of the operator
$$
(\widetilde{\mathcal L} \varphi)(x)= \int_{\obs} \frac{e^{-ik|x-y|}}{|x-y|} \varphi(y) dS_y, \quad x \in \partial \widetilde{\mathcal O}, \quad   \varphi\in L_2(\obs),
$$
is dense in $H^{3/2}(\partial \widetilde {\mathcal O})$. Then the same is true for $\overline{\widetilde{\mathcal L}}$. Hence for every $\varphi \in L_2(\obs)$, there exists a sequence $\psi_n \in L_2(\obs)$ such that $\overline{\widetilde{\mathcal L}}\psi_n\to \widetilde{\mathcal L}\varphi$ in $H^{3/2}(\partial \widetilde {\mathcal O})$. Below we consider functions $\overline{\widetilde{\mathcal L}}\psi_n, \widetilde{\mathcal L}\varphi, \overline{\mathcal L}\psi_n, \mathcal L\varphi$ defined by the corresponding integrals for all $x\in\mathbb R^3$. The standard a priory estimate (e.g., \cite{mcl}) for the solution $u=\overline{\widetilde{\mathcal L}}\psi_n- \widetilde{\mathcal L}\varphi$ of the Helmholtz equation in $\widetilde{O}$ implies that
$$
\|\overline{\widetilde{\mathcal L}}\psi_n- \widetilde{\mathcal L}\varphi\|_{H^{2}(\widetilde{\mathcal O})} \leq C(\lambda) \|\overline{\widetilde{\mathcal L}}\psi_n- \widetilde{\mathcal L}\varphi\|_{H^{3/2}(\partial \widetilde{\mathcal O})}\to 0  \quad  {\rm as} \quad n\to\infty.
$$
Since $\mathcal O \subset \widetilde{\mathcal O}$, we have that
$$
\|\mathcal L \varphi - \overline{\mathcal L} \psi_n \|_{H^{3/2}(\partial \mathcal O)}\to 0  \quad  {\rm as} \quad n\to\infty
$$
and
$$
\|\mathcal L \varphi - \overline{\mathcal L} \psi_n \|_{H^2( \mathcal O)}\to 0  \quad  {\rm as} \quad n\to\infty.
$$
The statement of the lemma is an immediate consequence of the last two relations and a priory estimates (e.g., \cite{mcl}) for the solutions of the problem (\ref{eq2}) (with radiation condition at infinity).

\qed

\begin{theorem}\label{ttt}
Consider two real-valued bounded potentials $n_1$ and $n_2$ and  their backscattering far-field operators $F_{S,i}, i=1,2$. If $\lambda=\lambda_0$ is not a Dirichlet eigenvalue for the domain $\mathcal S$, then the equality $F_{S,1}\varphi=F_{S,2}\varphi,~\lambda=\lambda_0,$ on a dense set $\{\varphi\}$ in $L_2(\obs)$ implies that  $\|n_1-n_2\|_{L^{\infty}}=0$.
\end{theorem}

The following lemma will be needed to prove the theorem above.

Denote by $F_0(\lambda),F^{out}(\lambda)$ Dirichlet-to-Neumann maps for the Helmholtz equation in the interior and exterior of $ \mathcal O$, respectively. The solutions are assumed to satisfy the radiation condition when $F^{out}$ is defined. Let $F_n$ be the Dirichlet-to-Neumann map for the equation $(\Delta+\lambda n)u=0$ in $ \mathcal O$.  The normal vector in all the cases is assumed to be directed outside of $\mathcal O$. Each of the Dirichlet-to-Neumann operators introduced above is a pseudo-differential operator of the first order and can be considered as a bounded operator from a Sobolev space $H^s(\partial \mathcal O)$ into $H^{s-1}(\partial \mathcal O),~s\in \mathbb R.$
\begin{lemma}\label{lsc}
The near field operator $F_S$ has the following representation:
\begin{equation}\label{6DecC1}
F_S=\frac{1}{4\pi}\mathcal L^* (  F_0-F^{out}) (F_n - F^{out})^{-1}(F_0 -F_n)\mathcal L.
\end{equation}
\end{lemma}
{\bf Remark.} These formulas are direct analogues of the formulas for the scattering amplitude in the problem of scattering of the plane waves (see \cite[Th.2.3]{lv2014}  in the case of the transmission problem). The only difference is that a plane wave is defined by the direction $\omega$ of the incident wave, and $\obs$ is replaced by the unit sphere $S^2=\{\omega: |\omega|=1\}$ in this case. The operators $\mathcal L, \mathcal L^*$ are also slightly different in the case of the plane waves. In particular,
\begin{equation}\label{opL1}
\mathcal L:L_2(S^2) \rightarrow L_2(\partial \mathcal O), \quad \mathcal L \varphi(x)= \int_{S^2}e^{ik\omega\cdot x} \varphi(\omega) dS_\omega.
\end{equation}

{\bf Proof}. Let us prove (\ref{6DecC1}). Note that  $u^{inc}|_{\partial \mathcal O}=\mathcal L \varphi$. We will look for $u^{sc}$  outside of $\mathcal O$ in the form of the potential $u^{sc}=\mathcal L^*\mu$ with an unknown density $\mu$, i.e.,
\begin{equation}\label{nnn}
u^{sc}=\int_{\partial \mathcal O} \frac{e^{ik|x-y|}}{|x-y|} \mu(y) dS_y, \quad x\in \mathbb R^3\backslash\mathcal O.
\end{equation}
More over, function $\mu$ must be chosen in such a way that $u^{sc}$  allows an extension in $ \mathcal O$ that satisfies (\ref{eq2}).

Every solutions of the Schr\"{o}dinger equation with a bounded potential belongs to $H^2(\mathcal O')$ for any bounded domain $\mathcal O'$. Therefore functions $u^{sc},u^{inc}$ and their normal derivatives are well defined on $\partial \mathcal O$. We reduce the scattering problem (\ref{eq2}),(\ref{rc}) to the following equation on $\partial \mathcal O$ for unknown $\mu$:
\begin{equation}\label{10FevA}
F_n(u^{sc}|_{\partial \mathcal O}+u^{inc}|_{\partial \mathcal O}) =F^{out}( u^{sc}|_{\partial \mathcal O}) + F_0( u^{inc}|_{\partial \mathcal O}).
\end{equation}
This equation follows from the fact that $u^{sc}+u^{inc}$ satisfies (\ref{eq2}) in $\mathcal O$, and $u^{sc}, u^{inc}$ are solutions of the Helmholtz equation in $\mathbb R^3\backslash \mathcal O$ and $\mathcal O$, respectively.

We note that operator $F_n$ is symmetric, and the imaginary part of the quadratic form of operator $F^{out}$ coincides with the total cross section, and therefore is positive (see \cite[Lemma 2.1]{lv2014}). Thus, operator $F_n-F^{out}$ is invertible, and equation (\ref{10FevA}) implies that
\begin{equation}\label{LS0428}
u^{sc}|_{\partial \mathcal O}=(F_n-F^{out})^{-1}(F_0-F_n)(u^{inc}|_{\partial \mathcal O})=(F_n-F^{out})^{-1}(F_0-F_n)\mathcal L \varphi.
\end{equation}

From (\ref{nnn}) it follows that $\mu=\frac{1}{4\pi}(F_0-F^{out})(u^{sc}|_{\partial \mathcal O}))$. It remains only to substitute (\ref{LS0428}) for $u^{sc}$ in the latter equation for $\mu$ and note that $F_S\varphi=u^{sc}|_\obs=\mathcal L^*\mu.$

\qed

{\bf Proof of Theorem \ref{ttt}.} We will reduce the statement of the theorem to the Gelfand-Calderon problem, which is solved in \cite[Th.1]{novikov} when $d=3$ and in \cite[Th.2.1]{bukh} when $d=2$.
%We will consider an arbitrary smooth domain $\mathcal O$, such that both supports of $n_1$ and $n_2$ are inside of $\mathcal O$ and $\mathcal S$ is outside of $\mathcal O$. We need an additional property in order to avoid request that operator $\Delta+\lambda n_i$  with Dirichlet b.c. does not have nontrivial kernel.

%We request that $\obs$ is located not too close to $\mathcal O$ in the following sense:  homotetia of $\mathcal O$ with coefficient close to 1 are still contain supports of $n_i$ and does not contain points of $\obs$. It is easy to show that the requested property holds for homotetia with coefficient delta arbitrarily close to $1$.

We preserve notations $F_0,F^{out}$ for the Dirichlet-to-Neumann maps for the Helmholtz equation in the interior and exterior of $ \mathcal O$, respectively, and we denote by $F_{n_1}, F_{n_2}$ the Dirichlet-to-Neumann maps for the Schr\"{o}dinger equations in $\mathcal O$ with potentials $\lambda n_1$ and $\lambda n_2$, respectively.

Operators
\begin{equation}\label{opop}
(F_0-F^{out}) (F_{n_i} - F^{out})^{-1}(F_0 -F_{n_i}):~L_2(\partial \mathcal O)\to L_2(\partial \mathcal O),~~~i=1,2,
\end{equation}
are bounded (and also compact). Indeed, each of the Dirichlet-to-Neumann operators introduced above is a pseudo-differential operator of the first order (non-smoothness of the potential does not play any role here, since the support of the potential is strictly inside of the domain). Their full symbols were calculated in \cite[Sect.3]{lv2013}. From this calculation it follows that operator $F_0-F^{out}$ has order one, operator $(F_{n_i} - F^{out})^{-1}$  has order $-1$, and a couple of the first terms of the full symbol of operator $F_0 -F_{n_i}$ vanish, i.e., the latter operator is compact. Thus (\ref{opop}) is compact.

Assume that data (\ref{data15}) for $n_1$ and $n_2$ coincide on a dense set $\{\varphi\}$ in $L_2(\obs)$. Then from Lemma \ref{6DecD} it follows that operators (\ref{opop}) are equal. The first factor from the left in (\ref{opop}) is an invertible operator (see the justification of the transition from (\ref{10FevA}) to (\ref{LS0428})). Hence, the equality of operators in  (\ref{opop}) implies that
$$
(F_{n_1} - F^{out})^{-1}(F_0 -F_{n_1})=(F_{n_2} - F^{out})^{-1}(F_0 -F_{n_2})
$$
as operators in $L_2(\partial \mathcal O)$. Adding and subtracting $F^{out}$ in the right factors, we get
$$
(F_{n_1} - F^{out})^{-1}(F_0 -F^{out})=(F_{n_2} - F^{out})^{-1}(F_0 -F^{out})
$$
 as operators in $L_2(\partial \mathcal O)$.
Hence, operators
$$
(F_{n_1} - F^{out})^{-1},~(F_{n_2} - F^{out})^{-1}:H^{-1}(\partial \mathcal O)\to L_2(\partial \mathcal O),
$$
are equal, and therefore,
$$
F_{n_1} - F^{out},~F_{n_2} - F^{out}:L_2(\partial \mathcal O)\to H^{-1}(\partial \mathcal O)
$$
are equal. Thus
$$
F_{n_1}\varphi = F_{n_2}\varphi
$$
for every $\varphi\in L_2(\partial \mathcal O)$.

Now uniqueness follows from \cite{bukh},\cite{XXXXX}  if $d=2$ and \cite{novikov} if $d=3$.

\end{document}